\documentclass{article}

\usepackage{microtype}
\usepackage{graphicx}
\usepackage{subfigure}
\usepackage{booktabs} 
\usepackage{amsmath}
\usepackage{amssymb}

\usepackage{bm}
\usepackage{algorithm}
\usepackage{algpseudocode}
\usepackage{xcolor}
\usepackage{multirow}

\newcommand{\armSR}{SESR }
\newcommand{\armSRDOT}{SESR} 
\newcommand{\armNPU}{Arm Ethos-N78 }
\newcommand{\armNPUDOT}{Arm Ethos-N78}
\graphicspath{{figs_sesr/}}
\usepackage{hyperref}

\usepackage[accepted]{mlsys2022}

\mlsystitlerunning{Collapsible Linear Blocks for Super-Efficient Super Resolution}

\begin{document}

\twocolumn[
\mlsystitle{Collapsible Linear Blocks for \\Super-Efficient Super Resolution}



\mlsyssetsymbol{equal}{*}

\begin{mlsysauthorlist}
\mlsysauthor{Kartikeya Bhardwaj}{arm}
\mlsysauthor{Milos Milosavljevic}{amazon}
\mlsysauthor{Liam O'Neil}{arm}
\mlsysauthor{Dibakar Gope}{armR}
\mlsysauthor{Ramon Matas}{armR}\\
\mlsysauthor{Alex Chalfin}{arm}
\mlsysauthor{Naveen Suda}{meta}
\mlsysauthor{Lingchuan Meng}{arm}
\mlsysauthor{Danny Loh}{arm}
\end{mlsysauthorlist}

\mlsysaffiliation{arm}{Arm Inc.}
\mlsysaffiliation{amazon}{Amazon}
\mlsysaffiliation{armR}{Arm Research}
\mlsysaffiliation{meta}{Meta}

\mlsyscorrespondingauthor{Kartikeya Bhardwaj}{kartikeya.bhardwaj@arm.com}
\mlsyscorrespondingauthor{Dibakar Gope}{dibakar.gope@arm.com}

\mlsyskeywords{Super Resolution, Mobile-Neural Processing Units, Efficient Inference}

\vskip 0.3in

\begin{abstract}
With the advent of smart devices that support 4K and 8K resolution, Single Image Super Resolution (SISR) has become an important computer vision problem. However, most super resolution deep networks are computationally very expensive. In this paper, we propose Super-Efficient Super Resolution (SESR) networks that establish a new state-of-the-art for efficient super resolution. Our approach is based on linear overparameterization of CNNs and creates an efficient model architecture for SISR. With theoretical analysis, we uncover the limitations of existing overparameterization methods and show how the proposed method alleviates them. Detailed experiments across six benchmark datasets demonstrate that \armSR achieves similar or better image quality than state-of-the-art models while requiring $\bm{2\times}$ to $\bm{330\times}$ fewer Multiply-Accumulate (MAC) operations. As a result, \armSR can be used on constrained hardware to perform $\times2$ (1080p to 4K) and $\times4$ (1080p to 8K) SISR. Towards this, we estimate hardware performance numbers for a commercial Arm mobile-Neural Processing Unit (NPU) for 1080p to 4K ($\times2$) and 1080p to 8K ($\times4$) SISR. Our results highlight the challenges faced by super resolution on AI accelerators and demonstrate that \armSR is significantly faster (e.g., $6\times$-$8\times$ higher FPS) than existing models on mobile-NPU. Finally, \armSR outperforms prior models by $1.5\times$-$2\times$ in latency on Arm CPU and GPU when deployed on a real mobile device. The code for this work is available at \url{https://github.com/ARM-software/sesr}.
\end{abstract}
]



\printAffiliationsAndNotice{}  
\begin{figure}[tb]
\centering
\includegraphics[width=0.46\textwidth]{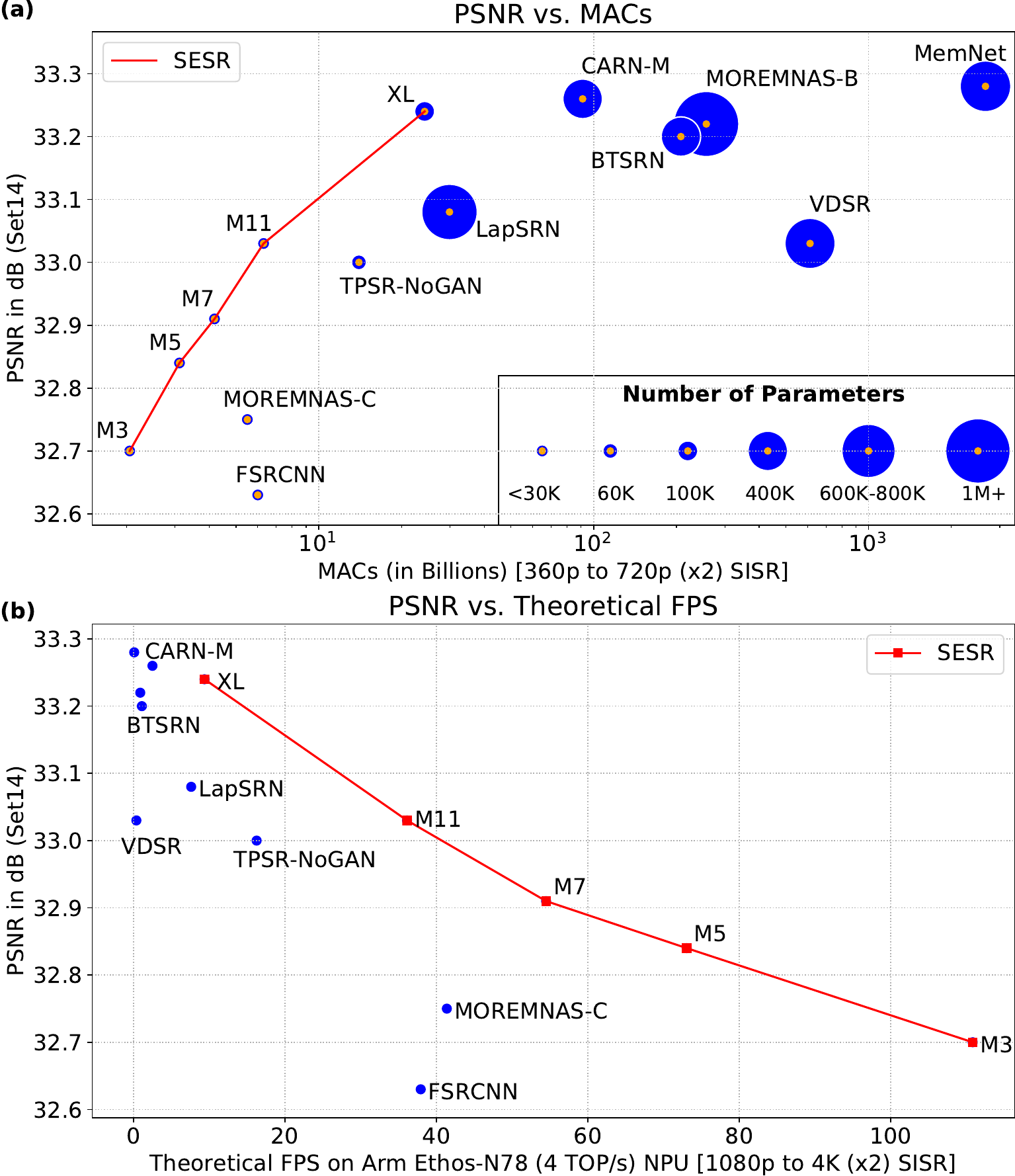}\vspace{-3mm}
	\caption{(a) PSNR on Set14 vs. MACs for different CNNs (360p to 720p, $\times 2$ SISR). (b) Most methods achieve less than 3FPS on a commercial \armNPU (4-TOP/s) mobile-NPU when performing 1080p to 4K SISR. \armSR establishes a new Pareto frontier for image quality-computation relationship.\vspace{-2mm}}
\label{fig1}
\end{figure}

\section{Introduction}\vspace{-2mm}
Single Image Super Resolution (SISR) is the classic ill-posed computer vision problem which aims to generate a high-resolution image from a low-resolution input. Recently, SISR and related super-sampling techniques have found applications in real-time upscaling of content up to 4K resolution~\cite{fbss,dlss2}. Moreover, with the advent of AI accelerators such as Neural Processing Units (NPUs) in upcoming 4K displays, laptops, and TVs~\cite{ethosN78}, 
AI-based upscaling of content to 4K resolution is now possible. Indeed, state-of-the-art SISR techniques are based on Convolutional Neural Networks (CNNs) which are computationally very expensive. Fig.~\ref{fig1}(a) shows the quality, as measured by PSNR, \textit{vs.}\ a measure of computational cost shown in SISR literature~\cite{carn,journeySR,tpsr,moremnas}, the Multiply-Accumulate (MAC) operations required to upscale an image from 360p to 720p. As evident, the existing models illustrate varied tradeoffs between image quality and computational costs. 

To put the published figures in context, consider a more realistic scenario of 1080p to 4K upscaling on a commercial \armNPUDOT, 4-Tera Ops per second (4-TOP/s)  mobile-NPU. This is an NPU suitable for deployment on smart devices such as smart phones, laptops, displays, TVs, etc.~\cite{ethosN78}.
Fig.~\ref{fig1}(b) shows the theoretical Frames Per Second (FPS) attained by various SISR networks. Clearly, even one of the smallest publicly available super resolution CNNs called FSRCNN~\cite{fsrcnn} can theoretically (the best case, $100\%$ hardware utilization scenario) achieve only 37 FPS on a 4-TOP/s NPU. When running on such constrained hardware, the larger deep networks are completely infeasible as most of them result in less than 3 FPS even in the best case. Hence, although many models like CARN-M~\cite{carn} have been designed to be lightweight, most SISR networks \textit{cannot} run on realistic, resource-constrained smart devices and mobile-NPUs. In addition, smaller models such as FSRCNN~\cite{fsrcnn} or TPSR~\cite{tpsr} do \textit{not} achieve high image quality. Therefore, there is a need for significantly smaller and much more accurate CNNs that attain high throughputs on resource-constrained devices.

To this end, we propose a new class of super resolution networks called \armSR that establish a new Pareto frontier on the quality-computation relationship (see Fig.~\ref{fig1}(a)). Driven by our insight that the challenge of on-device SISR is one of model training as much as of model architecture, we introduce an innovation that modifies the training protocol without modifying the inference-time network architecture.  Specifically, we propose \textit{Collapsible Linear Blocks}, which are sequences of linear convolutional layers that can be analytically collapsed into single, narrow (in terms of input/output channels) convolutional layers at inference time. This approach falls under the scope of \textit{linear overparameterization}~\cite{ExpandNets2020,repvgg}. We theoretically and empirically show the benefits of our proposed blocks. Overall, our work results in Super-Efficient Super Resolution (SESR) networks that demonstrate state-of-the-art tradeoff between image quality and computational costs. Fig.~\ref{fig1}(b) shows the theoretical FPS achieved by \armSR on the \armNPU (4-TOP/s) NPU. Clearly, several \armSR CNNs theoretically achieve nearly 60 FPS or more when performing 1080p to 4K SISR on a 4-TOP/s mobile-NPU.

Overall, we make the following \textbf{key contributions}:
\begin{enumerate}
    \item We propose \armSRDOT, a new class of super-efficient super resolution networks that establish a new state-of-the-art for efficient SISR. Towards this, we propose Collapsible Linear Blocks to train these networks. Our contribution is in both linear overaparameterization and overall model architecture design.
    \item We theoretically analyze existing overparameterization methods and discover that one of the recent methods does \textit{not} induce any change in gradient update compared to a completely non-overparameterized network. That is, under certain conditions (e.g., if the network is not too deep), such overparameterization methods do not present any advantages over the corresponding non-overparameterized models. The proposed \armSR fixes these limitations and improves gradient properties.
    \item Our results clearly demonstrate the superiority of \armSR over state-of-the-art models across six benchmark datasets for both $\times2$ and $\times4$ SISR. We achieve similar or better PSNR/SSIM than existing models while using $\bm{2\times}$ to $\bm{330\times}$ fewer MACs. Hence, \armSR can be used on constrained hardware to perform $\times2$ (1080p to 4K) and $\times4$ (1080p to 8K) SISR. We also present empirical evidence to support our theoretical insights. Moreover, we add SESR into a preliminary Neural Architecture Search (NAS) algorithm to improve the results further.
    \item Finally, we estimate hardware performance numbers for a commercial \armNPU NPU using its performance estimator for 1080p to 4K ($\times2$) and 1080p to 8K ($\times4$) SISR. These results clearly show the real-world challenges faced by SISR on AI accelerators and demonstrate that \armSR is substantially faster than existing models. We also discuss optimizations that can yield up to $8\times$ better runtime for 1080p to 4K SISR. When deployed on a real mobile device, \armSR is $1.5\times$-$2\times$ faster than baseline on Arm CPU and GPU.
\end{enumerate}

The rest of the paper is organized as follows: Section~\ref{sec:relWork} discusses the related work, while Section~\ref{sec:approach} describes our proposed approach. Section~\ref{sec:theory} presents theoretical insights behind \armSRDOT. Then, Section~\ref{sec:exp} demonstrates the effectiveness of \armSR over prior art and also presents hardware performance on mobile NPU, CPU, and GPU. Finally, Section~\ref{sec:conclusion} concludes the paper.

\section{Related Work}\label{sec:relWork}\vspace{-2mm}

\noindent
\textbf{Efficient SISR model design. }
While many excellent SISR methods have been proposed recently~\cite{DeepConvCVPR2016,MemNetICCV2017,ResDenseCVPR2018,ZhangECCV2018ChannelAttn,ZhangECCV2018ChannelAttn,LatticeNetECCV2020,WangACCV2020,Muqeet2020,zhao2020PixelAttn}, these are difficult to deploy on resource-constrained devices due to their heavy computational cost.
To this end, FSRCNN~\cite{fsrcnn} is a compact CNN for SISR. DRCN~\cite{drcncvpr2016} and DRRN~\cite{drrncvpr2017} adopt recursive layers to build deep network with fewer parameters. CARN~\cite{carn}, SplitSR~\cite{liu2021splitsr}, and
GhostSR~\cite{nie2021ghostsr} reduce the compute complexity by combining lightweight residual blocks with variants of group convolution. Since these and other model compression-based methods like~\cite{srdistillcvpr2018} are orthogonal to our linear overparameterization-based compact network design, they can be used in conjunction with \armSR to further reduce compute cost and model size.\vspace{1.4mm}

\noindent
\textbf{Perceptual SISR networks. }
Another set of SISR methods innovate towards novel perceptual loss functions and Generative Adversarial Networks (GANs)~\cite{srgan,esrgan,tpsr}. These techniques result in photo-realistic image quality. However, since our primary goal is to improve compute-efficiency, we only use traditional losses like Mean Absolute Error in this work.\vspace{1.4mm}
\begin{figure*}[tb]
\centering
\includegraphics[width=0.85\textwidth]{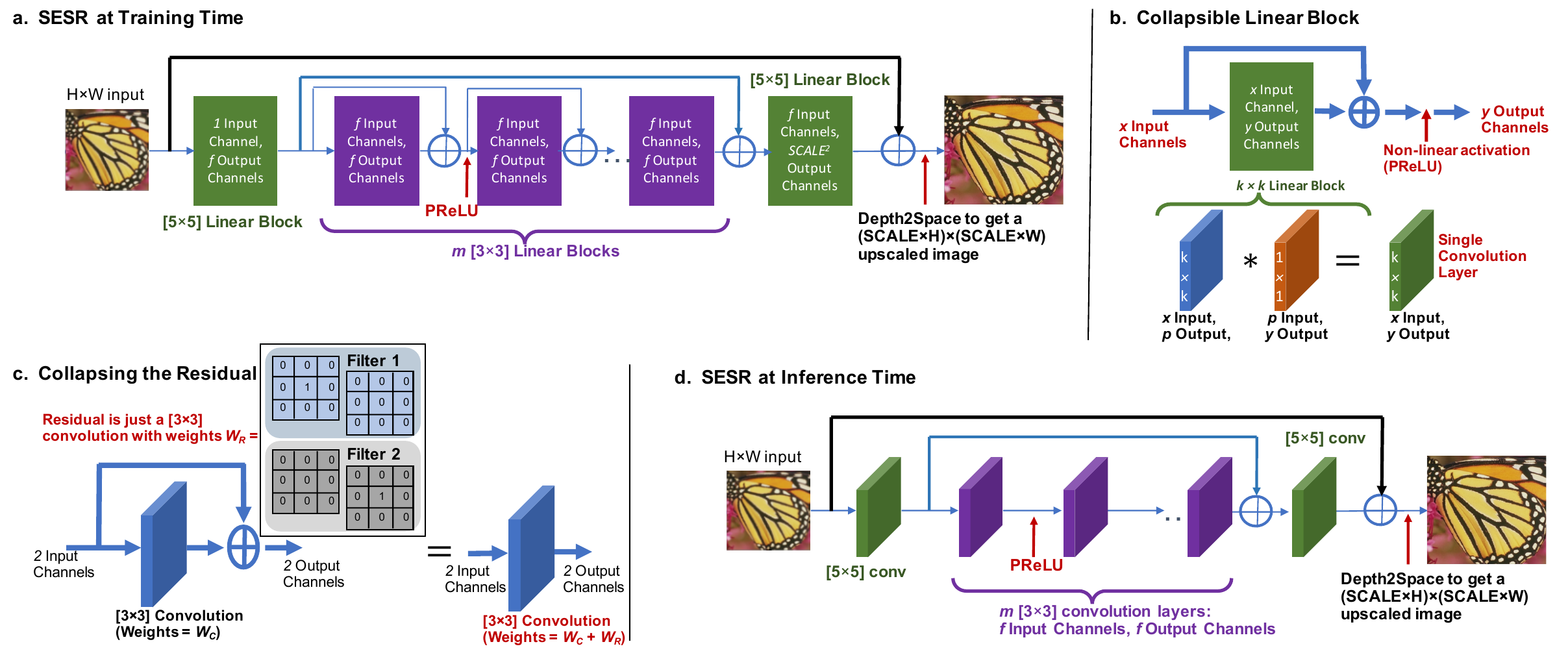}\vspace{-4mm}
	\caption{(a)~Proposed \armSR at training time contains two $5\times 5$ and $m$ $3\times 3$ linear blocks. There are two long residuals and several short residuals over $3\times 3$ linear blocks. (b)~A $k\times k$ linear block first uses a $k\times k$ convolution to project $x$ input channels to $p$ intermediate channels, which are projected back to $y$ output channels via a $1\times1$ convolution. (c)~Short residuals can further be collapsed into convolutions. (d)~Final inference time \armSR just contains two long residuals and $m$+$2$ narrow convolutions, resulting in a VGG-like CNN.\vspace{-3mm}}
\label{fig2}
\end{figure*}

\vspace{1mm}
\noindent
\textbf{Linear overparameterization in deep networks. }
There has been limited but important research on linear overparameterization~\cite{AroraICML2018,GCNICML2019,ACNetICCV2019, ExpandNets2020, repvgg} that shows the benefit of linearly overparameterized layers in speeding up the training of deep neural networks. 
Specifically, \cite{AroraICML2018} theoretically demonstrates that the linear overparameterization of fully connected layers can accelerate the training of deep linear networks by acting as a time-varying momentum and adaptive learning rate. Recent work on ExpandNets~\cite{ExpandNets2020} and ACNet~\cite{ACNetICCV2019} propose to overparameterize a convolutional layer and show that it accelerates the training of various CNNs and boosts the accuracy of the converged models. More recently, RepVGG~\cite{repvgg} proposed another overparameterization scheme that folds residual connections analytically so the final network looks like VGG.

Our approach differs from existing linear overparameterization works~\cite{ExpandNets2020, ACNetICCV2019, repvgg} in several ways: (\textit{i})~Linear overparameterization blocks have not been proposed for the super resolution problem; (\textit{ii})~We theoretically study the properties of various overparameterization methods: We found that ExpandNets run into vanishing gradient problems if not properly augmented with residual connections, and RepVGG gradient update is exactly the same as that for VGG. That is, for shallow networks, RepVGG and VGG will perform similarly. Our proposed method resolves these theoretical limitations of both the above methods; (\textit{iii})~We provide concrete empirical evidence towards our theoretical insights and demonstrate that our method is superior to ExpandNets and RepVGG; (\textit{iv})~Finally, existing methods do not design entirely new networks but rather augment existing networks like MobileNets~\cite{mobilenetv2} with overparameterized layers. At inference time, the collapsed network is the same as the original MobileNet. In contrast, \armSR innovates in both the linear block design as well as the overall inference model architecture to achieve state-of-the-art results for SISR.

\vspace{1.4mm}
\noindent
\textbf{NAS for lightweight super resolution. } 
NAS techniques have been shown to outperform manually designed networks in many applications~\cite{nasnet}. Therefore, recent NAS works target SISR by exploiting lightweight convolutional blocks such as group convolution, inverted residual blocks with different channel counts and kernel sizes, dilations, residual connections, upsampling layers, etc.~\cite{FALSR2019,Guo2020HierNAS,Song2020ResDense,wu2021trilevel,tpsr}. While our focus is not on NAS, we demonstrate that manually designed \armSR significantly outperforms existing state-of-the-art, NAS-designed SISR models. We also demonstrate that including the SESR blocks in a generic NAS further improves results over our method.

\vspace{-2mm}
\section{Super-Efficient Super Resolution}\label{sec:approach}\vspace{-2mm}
In this section, we explain the \armSR model architecture, collapsible linear blocks, and the inference \armSR network. We also present an efficient training methodology for SESR and how the proposed blocks can be used with NAS. 

\subsection{\armSR and Collapsible Linear Blocks}
Fig.~\ref{fig2}(a) illustrates \armSR network at training time. As evident, \armSR consists of multiple Collapsible Linear Blocks and several long and short residual connections. The structure of a linear block is shown in Fig.~\ref{fig2}(b). Essentially, a $k\times k$ linear block with $x$ input channels and $y$ output channels first expands activations to $p$ intermediate channels using a $k\times k$ convolution ($p>>x$). Then, a $1\times 1$ convolution is used to project the $p$ intermediate channels to $y$ final output channels. Since no non-linearity is used between these two convolutions, they can be \textit{analytically} collapsed into a single narrow convolution layer at inference time, hence, the name Collapsible Linear Blocks. The final collapsed convolution has $k\times k$ kernel size while using only $x$ input channels and $y$ output channels. Therefore, at training time, we train a very large deep network which gets analytically collapsed into a highly efficient deep network at inference time. This simple yet powerful overparameterization method, combined with residuals, shows significant benefits in convergence and image quality for SISR tasks.

We now describe the \armSR model architecture in detail (see Fig.~\ref{fig2}(a)). First, a $5\times5$ linear block is used to extract initial features from the input image. Next, the output of the first linear block passes through $m$ $3\times3$ linear blocks with \textit{short residuals}. Note that, a non-linearity (e.g., a Parameteric ReLU or PReLU) is used after this short residual addition and not before (see Fig.~\ref{fig2}(b)). The output of the first linear block is then added to the output of $m$ $3\times3$ linear blocks (see \textit{blue long-range residual} in Fig.~\ref{fig2}(a)). Following this, we use another $5\times5$ linear block to output SCALE$^2$ channels. At this point, the input image is added back to all output activations (see \textit{black long-range residual} in Fig.~\ref{fig2}(a)). Finally, a \textit{depth-to-space} operation converts the $H\times W\times \text{SCALE}^2$ activations into a ($\text{SCALE}\times H$)$\times$($\text{SCALE}\times W$) upscaled image. The depth-to-space operation described above is the same as the pixel shuffle part used inside subpixel convolutions~\cite{espcn, tpsr} and is one of the most standard techniques in SISR to obtain the upscaled images. Overall, our model is parameterized by $\{f, m\}$, where $f$ represents the number of output channels at all the linear blocks except the last one, and $m$ denotes the number of $3\times 3$ linear blocks used in the \armSR network. 

Note that, a single $k\times k$ convolution decomposed into a large $k\times k$ and a $1\times1$ convolution was used in ExpandNets~\cite{ExpandNets2020}. In Section~\ref{sec:theory}, we describe how this method without relevant residuals will result in vanishing gradient problems. We show empirical evidence towards this issue in ExpandNets in Section~\ref{sec:comparisonOver}. Hence, short residuals over the $3\times 3$ linear blocks are essential for good accuracy.

\vspace{-3mm}
\paragraph{Collapsing the Linear Block.} Once the \armSR network is trained, we can collapse the linear blocks into single convolution layers. Algorithm~\ref{alg1} shows the procedure to collapse the linear blocks which uses the following arguments: (\textit{i}) Trained weights ($W_{1:L}$) for all layers within the linear block, (\textit{ii}) Kernel Size ($k$) of linear block, (\textit{iii}) \textit{\#}Input channels ($N_{in}$), and (\textit{iv}) \textit{\#}Output channels ($N_{out}$). The output is the \textit{analytically} collapsed weight $W_C$ that replaces the linear block with a single small convolution layer.
\begin{algorithm}[!t]
{\small
    \caption{Collapse Linear Block}
    \label{alg1}
    \begin{algorithmic}[1] 
        \Procedure{collapse\_LB}{$W_{1:L}, k, N_{in}, N_{out}$}\\
            {\footnotesize \ \ \ \ \ \ \ \ \# First get NHWC tensor which will give the collapsed weight}
            \State $\Delta\gets \text{IDENTITY}(N_{in})$
            \State $\Delta\gets \text{expand\_dim}(\text{expand\_dim}(\Delta,1),1)$
            \State $\Delta\gets\text{ZERO\_PAD}(\Delta,[k-1,k-1])$
            
            \For{$i=1:L\ $} \Comment{{\footnotesize Go through all layers in Linear Block}}
                \If {$i == 1$}
                    \State $x\gets\text{Conv2D}(\Delta, W_i)$
                \Else
                    \State $x\gets\text{Conv2D}(x, W_i)$
                \EndIf
            \EndFor\label{euclidendwhile}
            \State $W_C \gets \text{transpose}(\text{reverse}(x,[1,2]),[1,2,0,3])$
            \State \textbf{return} $W_C$\Comment{{\footnotesize $W_C$ is the collapsed weight}}
        \EndProcedure
    \end{algorithmic}
}
\end{algorithm}
\begin{algorithm}[!t]
{\small
    \caption{Collapse Residual Addition into Convolution}
    \label{alg2}
    \begin{algorithmic}[1] 
        \Procedure{collapse\_residual}{$W_C$}
            \State $\text{shape}\gets W_C.\text{shape}$
            \State $\text{outChannels}, k\gets \text{shape}[3], \text{shape}[0]$
            \State $W_R\gets\text{ZEROS}(\text{shape})$
            \If {$k == 3$}
                \State $\text{idx}\gets 1$
            \EndIf
            \If {$k == 5$}
                \State $\text{idx}\gets 2$
            \EndIf
            \For{$i=1:\text{outChannels}\ $}
                \State $W_R[\text{idx},\text{idx},i,i]\gets 1$
            \EndFor
            \State \textbf{return} $W_R$\Comment{{\footnotesize $W_R$ is the residual weight}}
        \EndProcedure
    \end{algorithmic}
}
\end{algorithm}

\vspace{-3mm}
\paragraph{Collapsing the Residual into Convolutions.} Recall that, for our $3\times 3$ linear blocks, we perform a non-linearity after the residual additions. This allows us to collapse the residuals into collapsed convolution weights $W_C$. Fig.~\ref{fig2}(c) illustrates this process. Essentially, a residual is a $3\times 3$ convolution with identity weights, \textit{i.e.}, the output of this convolution is the same as its input. Fig.~\ref{fig2}(c) shows what this weight looks like for a residual add with two input and output channels. Algorithm~\ref{alg2} shows a concrete pseudo code for collapsing the residual into a convolution. The final single convolution weight (combining both linear block and residual) is then given by $W_{3\times3} = W_C + W_R$.

\vspace{-3mm}
\paragraph{Why not use standard ResNet skip connections?}
If regular residual connections are used like in ResNets (which cannot be collapsed), the resulting model will not be suitable for constrained hardware since the residual connections require additional memory transactions which can be very expensive for SISR tasks since the feature maps are very large (e.g., $1920\times 1080$, etc.). This is why, collapsing the residuals using Algorithm~\ref{alg2} is very important.

\subsection{\armSR at Inference Time}
The final, inference time \armSR network architecture is shown in Fig.~\ref{fig2}(d). As evident, all linear blocks and short residuals are collapsed into single convolutions. Hence, the final inference network is nearly a VGG-like CNN: Just $m+2$ convolution layers with most having $f$ output channels, and two additional long residuals (see blue and black residuals in Fig.~\ref{fig2}(d)). For this network, \textit{\#}parameters for $\times2$ SISR is given by $P = (5\times5\times1\times f)+m\times(3\times3\times f\times f) + (5\times5\times f\times 4)$\footnote{Following standard practice~\cite{fsrcnn}, we convert the RGB image into Y-Cb-Cr and use only the Y-channel for super resolution. Thus, \armSR has only one input and one output channel.}. Then, \textit{\#}MACs can be calculated as \textit{\#}MACs $ = H\times W\times P$, where $H, W$ are the dimensions of the low resolution input. We obtain the best PSNR results using the inference network architecture shown in Fig.~\ref{fig2}(d). However, to achieve even better hardware efficiency, we create another version of \armSR that removes the long black residual and replaces PReLU with ReLU. We found that this has a minimal impact on image quality (detailed ablations in Section~\ref{sec:preluAbl}).

\subsection{Efficient Training Methodology}\label{sec:etm}
\begin{figure}[tb]
\centering
\includegraphics[width=0.48\textwidth]{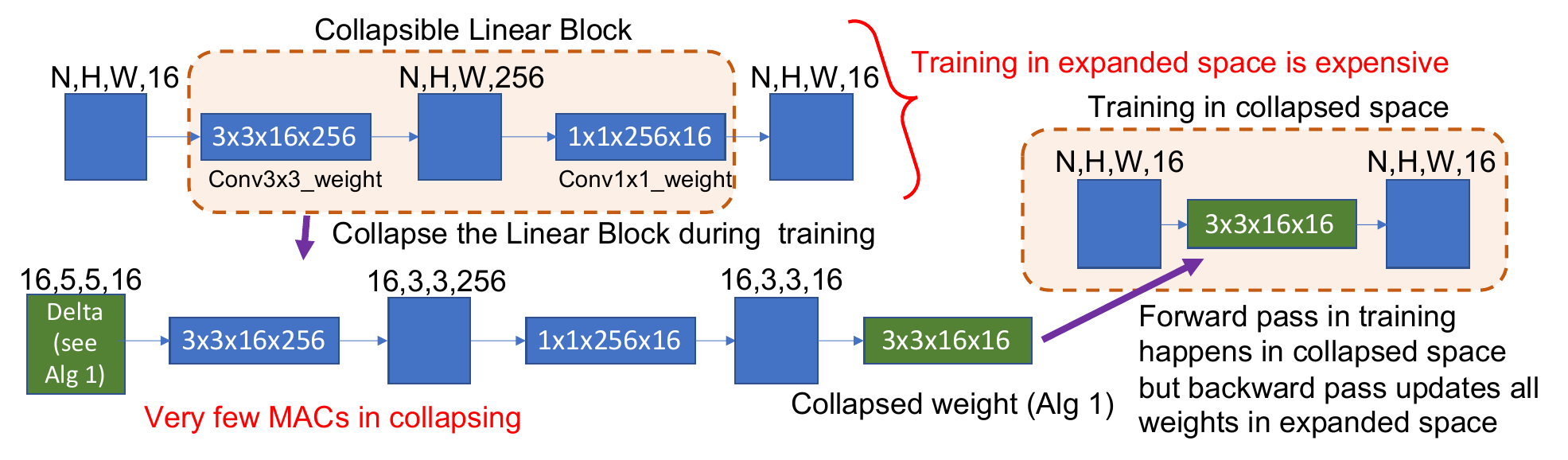}\vspace{-4mm}
	\caption{Collapsing (while training) is very efficient since the image size is 5$\times$5 and batch size (N) is 16. These are much smaller than N, H, W when training in expanded space (i.e., N=32, image size is H$\times$W = $64\times 64$). Example: in expanded space, we have 32,64,64,256 tensors on which last 1$\times$1 conv operates. In our efficient implementation, 1$\times$1 conv operates on 16,3,3,256 tensor.\vspace{-3mm}}
\label{effTr}
\end{figure}

The training time would increase if we directly train collapsible linear blocks in the expanded space and collapse them later. To address this, we developed an efficient implementation of \armSRDOT: We collapse the Linear Blocks at each training step (using Algorithms~\ref{alg1} and~\ref{alg2}), and then use this collapsed weight to perform forward pass convolutions. Since model weights are very small tensors compared to feature maps, this collapsing takes a very small time. \textit{The training (backward pass) still updates the weights in the expanded space but the forward pass happens in collapsed space even during training} (see Fig.~\ref{effTr}). 
Specifically, for the \armSRDOT-M5 and a batch of 32 [64$\times$64] images, training in expanded space takes 41.77B MACs for a single forward pass, whereas our efficient implementation takes only 1.84B MACs. Similar improvements happen in GPU memory and backward pass (due to reduced size of layerwise Jacobians). Therefore, training \armSR networks is highly efficient.

\subsection{SESR with Even-Sized and Asymmetric Kernels}
While convolutions with even-sized kernels~\cite{shuang_neurips2019} and asymmetric kernels~\cite{ACNetICCV2019} have been explored in the recent past, there has not been any work yet to demonstrate their true potential for better performance on commercial NPUs. Use of smaller even-sized (e.g., $2\times2$ kernels) and asymmetric kernels (e.g., $3\times2$ kernels) in place of $3\times3$ kernels requires fewer operations and less memory to perform convolution operations. Exploiting this insight, we employ a generic differentiable NAS (DNAS) with appropriate constraints to search for SESR networks that can accommodate collapsible linear blocks potentially with smaller even-sized and asymmetric kernels to further reduce the computational complexity and improve the inference time without compromising accuracy. 

DNAS requires a backbone supernet to be defined as the starting point for the search. We use a SESR backbone, consisting of a series of collapsible linear blocks. Each collapsible linear block can choose the height and width of the convolution kernel. This promotes differently-sized kernels during NAS. A skip connection branch ($1\times1$ convolution if the parallel convolutional block downsamples the input) is also added in parallel to each collapsible linear block in the backbone to create shortcuts for choosing the number of layers. We use DNAS to choose the size of the kernels, the number of channels, and the number of layers in this backbone network while trying to satisfy the hardware constraints. Since the inference time plays an important role in model design, we incorporate a latency constraint into our DNAS (following standard hardware-aware DNAS practices). Note that, \textit{this is only a preliminary proof-of-concept to show that introducing SESR into NAS further improves results over our manually designed network}. The main focus of this work is on manual architecture design and linear overparameterization and not on NAS.

\section{Theoretical Properties of \armSRDOT}\label{sec:theory}
\begin{figure}[tb]
\centering
\includegraphics[width=0.48\textwidth]{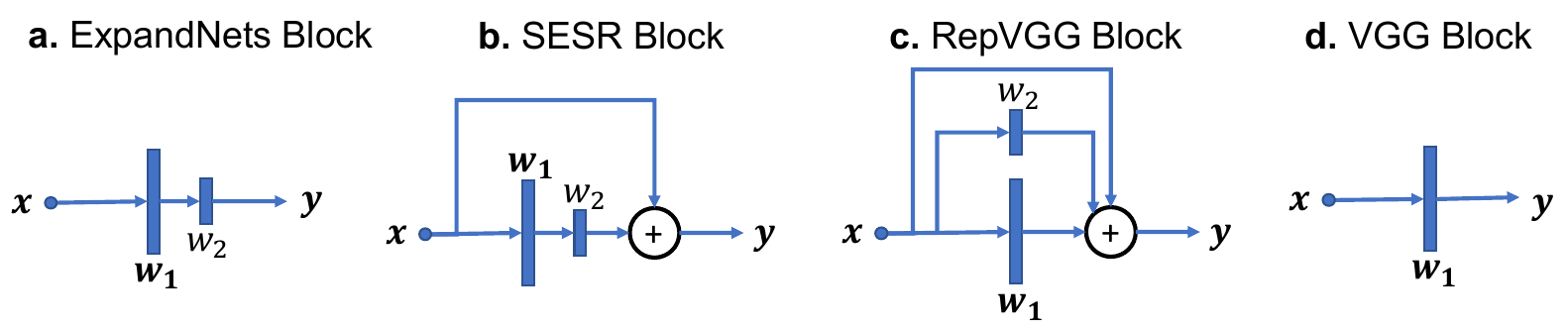}\vspace{-1mm}\vspace{-3mm}
	\caption{Different types of overparameterization schemes: (a)~ExpandNets, (b) SESR, (c) RepVGG, and (d) VGG blocks. Blocks shown in (a,b,c) collapse into the VGG block (d).\vspace{-3mm}}
\label{blocks}
\end{figure}

\vspace{-1mm}
We now study the theoretical properties of the collapsible linear block proposed above and compare it to existing overparameterization schemes: (\textit{i})~ExpandNets~\cite{ExpandNets2020, AroraICML2018}, and (\textit{ii})~RepVGG~\cite{repvgg}. In the following subsections, we assume linear overparameterized networks~\cite{AroraICML2018} where all blocks have the original linear layer weights as $\bm{w_1}$. Without loss of generality and following~\cite{AroraICML2018}, the layers are overparameterized by a \textit{single scalar} $w_2$ parameter. We denote the collapsed weight as $\bm{\beta}$. Since we need to study impact of short residuals explicitly, we assume that input and output have same dimensions. Various overparameterization schemes and their differences are depicted in Fig.~\ref{blocks}.
Consider a standard $l_2$ loss linear regression problem:
\begin{equation}
    \mathcal{L(\beta)} = \mathbb{E}_{\bm{x,y}\sim\mathcal{D}}\left[\frac{1}{2}||\bm{x}^T\bm{\beta}-\bm{y}||_2^2\right],
\label{lploss}
\end{equation}
where, $\bm{x}$ is input data, $\bm{y}$ is the output, and $\mathcal{D}$ is the training dataset. 
The gradient for the collapsed weight $\bm{\beta}$ is:
\begin{equation}
    \nabla_{\bm{\beta}} = \mathbb{E}_{\bm{x,y}\sim\mathcal{D}}\left[(\bm{x}^T\bm{\beta}-\bm{y})\bm{x}\right].
\label{gradBeta}
\end{equation}
We next compute the gradient update rules for different kinds of overparameterization blocks to understand the limitations of existing blocks and how the proposed method alleviates them.

\subsection{Gradient update for ExpandNets}\label{sec:theoryExp}\vspace{-1mm}
We briefly review the gradient update rule for ExpandNets kind of linear overparameterization as derived in~\cite{AroraICML2018}. The collapsed weight for ExpandNets is given by $\bm{\beta}=\bm{w_1}w_2$ (see Fig.~\ref{blocks}(a)). Then, by chain rule, we get $\nabla_{\bm{w_1}} = \nabla_{\bm{\beta}}w_2$. Thus, the gradient update is given by:
\begin{equation}
{\small
\begin{aligned}
    \bm{\beta}^{(t+1)} &= (\bm{w_1}^{(t+1)})({w_2}^{(t+1)})\\
    &= (\bm{w_1}^{(t)}-\eta\nabla_{{\bm{w_1}}^{(t)}})({w_2}^{(t)}-\eta\nabla_{{w_2}^{(t)}})\\
    &= \bm{w_1}^{(t)}{w_2}^{(t)} - \eta\nabla_{{\bm{w_1}}^{(t)}}{w_2}^{(t)} - \eta\nabla_{{w_2}^{(t)}}\bm{w_1}^{(t)} + \mathcal{O}(\eta^2)\\
    &= \bm{\beta}^{(t)} - \eta({{w_2}^{(t)}})^2\nabla_{\bm{\beta}^{(t)}} - \eta\nabla_{{w_2}^{(t)}}{({w_2}^{(t)}})^{-1}\bm{\beta}^{(t)}\\
    &= \bm{\beta}^{(t)} - \rho^{(t)}\nabla_{\bm{\beta}^{(t)}} - \gamma^{(t)}\bm{\beta}^{(t)}.
\end{aligned}
}
\label{expUpdate}
\end{equation}
Here, $\mathcal{O}(\eta^2)$ terms have been ignored since the learning rate $\eta$ is small. Arora et al. explain that linear overparameterization results in time varying momentum and adaptive learning rate and is often better (empirically) than optimization strategies like AdaDelta and ADAM~\cite{AroraICML2018}.

\subsection{Gradient update for \armSRDOT}\label{sec:theoryClb}\vspace{-1mm}
Due to the identity connection, the collapsed weight for \armSR is given as: $\bm{\beta}=\bm{w_1}w_2+\bm{I}$ (see Fig.~\ref{blocks}(b)). Similar to ExpandNets, we get $\nabla_{\bm{w_1}} = \nabla_{\bm{\beta}}w_2$. Therefore, following~\cite{AroraICML2018}, the update for collapsed weight in \armSR can be computed as:
\begin{equation}
{\small
\begin{aligned}
    \bm{\beta}^{(t+1)} &= (\bm{w_1}^{(t+1)})({w_2}^{(t+1)}) + \bm{I}\\
    &= (\bm{w_1}^{(t)}-\eta\nabla_{{\bm{w_1}}^{(t)}})({w_2}^{(t)}-\eta\nabla_{{w_2}^{(t)}}) + \bm{I}\\
    &= \bm{w_1}^{(t)}{w_2}^{(t)} - \eta\nabla_{{\bm{w_1}}^{(t)}}{w_2}^{(t)} - \eta\nabla_{{w_2}^{(t)}}\bm{w_1}^{(t)} + \bm{I} + \mathcal{O}(\eta^2)\\
    &= \bm{\beta}^{(t)} - \eta({{w_2}^{(t)}})^2\nabla_{\bm{\beta}^{(t)}} - \eta\nabla_{{w_2}^{(t)}}{({w_2}^{(t)}})^{-1}(\bm{\beta}^{(t)}-\bm{I})\\
    &= \bm{\beta}^{(t)} - \rho^{(t)}\nabla_{\bm{\beta}^{(t)}} - \gamma^{(t)}\bm{\beta}^{(t)} {\color{red} +\  \gamma^{(t)}},
\end{aligned}
}
\label{clbUpdate}
\end{equation}
where, $\rho^{(t)}=\eta({{w_2}^{(t)}})^2$ and $\gamma^{(t)}=\eta\nabla_{{w_2}^{(t)}}{({w_2}^{(t)}})^{-1}$. Therefore, like ExpandNets~\cite{AroraICML2018,ExpandNets2020}, \armSR update results in a time varying momentum ($\gamma^{(t)}$) and adaptive learning rate ($\rho^{(t)}$). However, \textit{the \armSR update is even more adaptive than ExpandNets since we get an extra $\gamma^{(t)}$ term in Eq.~\eqref{clbUpdate} due to the identity connection}. Moreover, as well established in literature, identity connection help with information propagation in deep networks and prevent vanishing gradients~\cite{resnet,nnmass}. Therefore, besides the more adaptive update in \armSRDOT, the short residuals also result in better gradient flow through the network and enhance its trainability. We will show concrete empirical results demonstrating this.

\subsection{Gradient update for RepVGG}\label{sec:theoryRep}\vspace{-1mm}
Since the recent RepVGG block also introduced overparameterization with the short residuals, a natural question is whether this block also results in more adaptive updates like our proposed \armSR block. The collapsed weight in RepVGG is given by: $\bm{\beta}=\bm{w_1} + w_2\bm{I}+\bm{I}$ (see Fig.~\ref{blocks}(c)). By chain rule, $\nabla_{\bm{w_1}} = \nabla_{w_2} = \nabla_{\bm{\beta}}$. Therefore, the gradient update for the collapsed weight is as follows:
\begin{equation}
{\small
\begin{aligned}
    \bm{\beta}^{(t+1)} &= \bm{w_1}^{(t+1)} + {w_2}^{(t+1)} + \bm{I}\\
    &= (\bm{w_1}^{(t)}-\eta\nabla_{{\bm{w_1}}^{(t)}}) + ({w_2}^{(t)}-\eta\nabla_{{w_2}^{(t)}}) + \bm{I}\\
    &= \bm{\beta}^{(t)} - 2\eta\nabla_{\bm{\beta}^{(t)}}\\
    &= \bm{\beta}^{(t)} - \lambda\nabla_{\bm{\beta}^{(t)}},
\end{aligned}
}
\label{rvggUpdate}
\end{equation}
where, $\lambda=2\eta$ is a constant. Therefore, RepVGG update does \textit{not} result in any adaptivity or time varying momentum or learning rates. In fact, \textit{the gradient update for RepVGG looks exactly like that for a VGG network} ($\bm{\beta}^{(t)}=\bm{w_1}^{(t)}$ for VGG; see Fig.~\ref{blocks}(d)). 

Practically, RepVGG would avoid the vanishing gradient problem due to the presence of short residuals for very deep networks. However, many networks for mobile and edge devices are often small and do not have that many layers. For example, our networks have a maximum of 13 layers when collapsed. The vanishing gradient problem will still appear for these networks if the convolutions are expanded using linear blocks (i.e., the network would have total 26 layers which can become harder to train if short residuals are not present). However, in case of RepVGG, the expanded network still has 13 layers which would not be hard to train. Therefore, for such shallow networks, RepVGG is expected to behave similar to VGG networks particularly because their gradient updates are equivalent. In Section~\ref{sec:comparisonOver}, we present empirical evidence which precisely shows that a 13 layer inference network trained with linear blocks without short residuals (i.e., total 26 layers due to an ExpandNet block) has poor trainability because of vanishing gradients. Also, a 13 layer inference model with RepVGG block -- trained with $k\times k$ convolutions, a $1\times1$ convolution branch, and short residuals -- does not improve over a model trained with completely collapsed structure (e.g., if we train the VGG-like model in Fig.~\ref{fig2}(d) directly without short residuals and linear blocks). In contrast, SESR with proposed blocks significantly outperforms ExpandNets and RepVGG.

\section{Experimental Setup and Results}\label{sec:exp}
We first describe our setup and results for \armSR on six datasets. We then quantify the improvements in training time as a result of our proposed efficient training method. We further compare \armSR with ExpandNets and RepVGG. Next, we estimate hardware performance for 1080p to 4K ($\times2$) and 1080p to 8K ($\times4$) SISR on \armNPU NPU and present our NAS results with SESR search space. We also present open source performance estimation for Arm Ethos-U55 NPU. Finally, we deploy \armSR on a real mobile device to quantify performance on Arm CPU and GPU.

\begin{table*}[!t]
\caption{PSNR/SSIM results on $\times2$ Super Resolution on several benchmark datasets. MACs are reported as the number of multiply-adds needed to convert an image to 720p ($1280\times 720$) resolution via $\times2$ SISR. \textcolor{red}{Red}/\textcolor{blue}{Blue} indicate \textcolor{red}{Best}/\textcolor{blue}{Second Best} within each regime.}
\centering
\scalebox{0.7}{
\begin{tabular}{|c|l|l|l|c|c|c|c|c|c|}
\hline
Regime & Model & Parameters & MACs & Set5 & Set14 & BSD100 & Urban100 & Manga109 & DIV2K \\ \hline \hline
\multirow{7}{*}{Small}&Bicubic     & $-$  & $-$  & 33.68/0.9307  &  30.24/0.8693  & 29.56/0.8439  & 26.88/0.8408  & 30.82/0.9349  & 32.45/0.9043   \\ \cline{2-10}
&FSRCNN \textbf{(our setup)}     & \textcolor{blue}{12.46K}  & 6.00G  & 36.85/0.9561 & 32.47/0.9076    &  31.37/0.8891 & 29.43/0.8963  & 35.81/0.9689  &  34.73/0.9349  \\ \cline{2-10}
&FSRCNN~\cite{fsrcnn}     & \textcolor{blue}{12.46K}  & 6.00G  & 36.98/0.9556  & 32.62/0.9087    &  31.50/0.8904 & 29.85/0.9009  & 36.62/0.9710  &  34.74/0.9340  \\ \cline{2-10}
&MOREMNAS-C~\cite{moremnas} & 25K & 5.5G & 37.06/0.9561 & 32.75/0.9094 & 31.50/0.8904 & 29.92/0.9023 & $-/-$ & $-/-$ \\ \cline{2-10} 
&\armSRDOT-M3 ($f$=16, $m$=3)    & \textcolor{red}{8.91K}  & \textcolor{red}{2.05G}  & 37.21/0.9577  &  32.70/0.9100  &  31.56/0.8920 & 29.92/0.9034  & 36.47/0.9717  &  35.03/0.9373 \\ \cline{2-10}
&\textcolor{blue}{\armSRDOT-M5 ($f$=16, $m$=5)}    & 13.52K  & \textcolor{blue}{3.11G}  & \textcolor{blue}{37.39/0.9585}  &  \textcolor{blue}{32.84/0.9115}  & \textcolor{blue}{31.70/0.8938}  & \textcolor{blue}{30.33/0.9087}  & \textcolor{blue}{37.07/0.9734}  &  \textcolor{blue}{35.24/0.9389}  \\ \cline{2-10}
&\textcolor{red}{\armSRDOT-M7 ($f$=16, $m$=7)}    & 18.12K  & 4.17G  & \textcolor{red}{37.47/0.9588} &  \textcolor{red}{32.91/0.9118}  & \textcolor{red}{31.77/0.8946}  & \textcolor{red}{30.49/0.9105}  & \textcolor{red}{37.14/0.9738}  & \textcolor{red}{35.32/0.9395}  \\ \hline\hline
\multirow{2}{*}{Medium}&\textcolor{blue}{TPSR-NoGAN~\cite{tpsr}} & \textcolor{blue}{60K} & \textcolor{blue}{14.0G} & \textcolor{blue}{37.38/0.9583} & \textcolor{blue}{33.00/0.9123} & \textcolor{blue}{31.75/0.8942} & \textcolor{blue}{30.61/0.9119} & $-/-$ & $-/-$ \\ \cline{2-10}
&\textcolor{red}{\armSRDOT-M11 ($f$=16, $m$=11)}    & \textcolor{red}{27.34K}  & \textcolor{red}{6.30G}  & \textcolor{red}{37.58/0.9593}  &  \textcolor{red}{33.03/0.9128}  & \textcolor{red}{31.85/0.8956}  & \textcolor{red}{30.72/0.9136}  & \textcolor{red}{37.40/0.9746} & \textcolor{red}{35.45/0.9404} \\ \hline\hline
\multirow{6}{*}{Large}&VDSR~\cite{vdsr}     & 665K  & 612.6G  & 37.53/0.9587  &   33.05/0.9127 & 31.90/0.8960  & 30.77/0.9141  & 37.16/0.9740  & \textcolor{blue}{35.43/0.9410}  \\ \cline{2-10}
&LapSRN~\cite{lapsrn} & 813K & \textcolor{blue}{29.9G} & 37.52/\textcolor{blue}{0.9590} & 33.08/0.9130 & 31.80/0.8950 & 30.41/0.9100 & \textcolor{blue}{37.53/0.9740}  & 35.31/0.9400  \\ \cline{2-10}
&BTSRN~\cite{btsrn} & \textcolor{blue}{410K} & 207.7G & \textcolor{blue}{37.75}/$-$ & 33.20/$-$ & \textcolor{red}{32.05}/$-$ & \textcolor{red}{31.63}/$-$ & $-/-$  & $-/-$  \\ \cline{2-10} 
&\textcolor{blue}{CARN-M~\cite{carn}}     & 412K  & 91.2G  & 37.53/0.9583  &   \textcolor{red}{33.26}/\textcolor{blue}{0.9141} & 31.92/\textcolor{blue}{0.8960}  & \textcolor{blue}{31.23}/\textcolor{red}{0.9193}  & $-/-$  & $-/-$  \\ \cline{2-10} 
&MOREMNAS-B~\cite{moremnas} &  1118K & 256.9G & 37.58/0.9584 & 33.22/0.9135 & 31.91/0.8959 & 31.14/0.9175 & $-/-$ & $-/-$ \\ \cline{2-10}
&\textcolor{red}{\armSRDOT-XL ($f$=32, $m$=11)}    & \textcolor{red}{105.37K}  & \textcolor{red}{24.27G}  & \textcolor{red}{37.77/0.9601}  &  \textcolor{blue}{33.24}/\textcolor{red}{0.9145}  & \textcolor{blue}{31.99}/\textcolor{red}{0.8976}  & 31.16/\textcolor{blue}{0.9184}  & \textcolor{red}{38.01/0.9759} & \textcolor{red}{35.67/0.9420} \\ 
\hline
\end{tabular}
\label{resX2}
}\vspace{-2mm}
\end{table*}

\subsection{Experimental Setup}\vspace{-1mm}
We train our \armSR networks for 300 epochs using ADAM optimizer with a constant learning rate of $5\times 10^{-4}$ and a batch size of 32 on DIV2K training set. We use mean absolute error ($l_1$) loss between the high resolution and generated images to train \armSRDOT. For training efficiency, we take 64 random crops of size $64\times64$ from each image; hence, each epoch conducts $800\times 64/32=1600$ training steps. We vary the number of $3\times3$ linear blocks ($m$) as $\{3,5,7,11\}$ and keep number of channels as $f=16$. We also train an extra-large model for \armSR (called \armSRDOT-XL), where $f=32$ and $m=11$. Also, we set the expanded number of channels within linear blocks (parameter $p$ in Fig.~\ref{fig2}(b)) as 256. The models are collapsed using Algorithms~\ref{alg1},~\ref{alg2} and are tested on six standard SISR datasets: Set5, Set14, BSD100, Urban100, Manga109, and DIV2K validation set. Following standard practice, only Y-channel is used to compute PSNR/SSIM. 

For $\times4$ SISR, we start with the pretrained $\times2$ \armSR networks. We first replace the final $5\times5\times f\times4$ layer by $5\times 5\times f\times 16$ and then perform the depth-to-space operation twice. Note that, this is different from many prior SISR networks which repeat the upsampling block (containing a convolution \textit{and} a depth-to-space operation) multiple times~\cite{tpsr}. In contrast, we do a single convolution and apply depth-to-space twice. This helps us save additional MACs for $\times4$ SISR. We will elaborate on this in the results section. \armSR is implemented in TensorFlow and trained on a single NVIDIA V100 GPU.
\begin{table*}[!t]
\caption{PSNR/SSIM results on $\times4$ Super Resolution on several benchmark datasets. MACs are reported as the number of multiply-adds needed to convert an image to 720p ($1280\times 720$) resolution via $\times4$ SISR. \textcolor{red}{Red}/\textcolor{blue}{Blue} indicate \textcolor{red}{Best}/\textcolor{blue}{Second Best} within each regime.}
\centering
\scalebox{0.7}{
\begin{tabular}{|c|l|l|l|c|c|c|c|c|c|}
\hline
Regime & Model & Parameters & MACs & Set5 & Set14 & BSD100 & Urban100 & Manga109 & DIV2K \\ \hline \hline
\multirow{6}{*}{Small} & Bicubic     & $-$  & $-$  & 28.43/0.8113  &  26.00/0.7025  & 25.96/0.6682  & 23.14/0.6577  & 24.90/0.7855  & 28.10/0.7745   \\ \cline{2-10}
& FSRCNN \textbf{(our setup)}     & \textcolor{red}{12.46K}  & 4.63G  & 30.45/0.8648 & 27.44/0.7528    &  26.89/0.7124 & 24.39/0.7212  & 27.40/0.8539  &  29.37/0.8117  \\ \cline{2-10} 
& FSRCNN~\cite{fsrcnn}     & \textcolor{red}{12.46K}  & 4.63G  & 30.70/0.8657 & 27.59/0.7535 & 26.96/0.7128 & 24.60/0.7258 & 27.89/0.8590 & 29.36/0.8110 \\ \cline{2-10}
& \armSRDOT-M3 ($f$=16, $m$=3)    & \textcolor{blue}{13.71K}  & \textcolor{red}{0.79G}  & 30.75/0.8714  &  27.62/0.7579  &  27.00/0.7166 & 24.61/0.7304  & 27.90/0.8644  &  29.52/0.8155 \\ \cline{2-10}
& \textcolor{blue}{\armSRDOT-M5 ($f$=16, $m$=5)}    & 18.32K  & \textcolor{blue}{1.05G}  & \textcolor{blue}{30.99/0.8764}  &  \textcolor{blue}{27.81/0.7624}  & \textcolor{blue}{27.11/0.7199}  & \textcolor{blue}{24.80/0.7389}  & \textcolor{blue}{28.29/0.8734}  &  \textcolor{blue}{29.65/0.8189}  \\ \cline{2-10}
& \textcolor{red}{\armSRDOT-M7 ($f$=16, $m$=7)}    & 22.92K  & 1.32G  & \textcolor{red}{31.14/0.8787} &  \textcolor{red}{27.88/0.7641}  & \textcolor{red}{27.13/0.7209}  & \textcolor{red}{24.90/0.7436}  & \textcolor{red}{28.53/0.8778}  & \textcolor{red}{29.72/0.8204}  \\ \hline \hline
\multirow{2}{*}{Medium} & \textcolor{blue}{TPSR-NoGAN~\cite{tpsr}} & \textcolor{blue}{61K} & \textcolor{blue}{3.6G} & \textcolor{blue}{31.10/0.8779} & \textcolor{red}{27.95/0.7663} & \textcolor{blue}{27.15/0.7214} & \textcolor{blue}{24.97/0.7456} & $-/-$ & $-/-$ \\ \cline{2-10}
& \textcolor{red}{\armSRDOT-M11 ($f$=16, $m$=11)}    & \textcolor{red}{32.14K}  & \textcolor{red}{1.85G}  & \textcolor{red}{31.27/0.8810}  &  \textcolor{blue}{27.94/0.7660}  & \textcolor{red}{27.20/0.7225}  & \textcolor{red}{25.00/0.7466}  & \textcolor{red}{28.73/0.8815} & \textcolor{red}{29.81/0.8221} \\ \hline \hline
\multirow{5}{*}{Large} & VDSR~\cite{vdsr}     & 665K  & 612.6G  & 31.35/0.8838 & 28.02/0.7678 & 27.29/0.7252 & 25.18/0.7525 & 28.82/0.8860 & 29.82/0.8240 \\ \cline{2-10}
& LapSRN~\cite{lapsrn} & 813K & 149.4G &  31.54/0.8850 & 28.19/\textcolor{blue}{0.7720} & 27.32/\textcolor{blue}{0.7280} & 25.21/0.7560 & \textcolor{red}{29.09}/\textcolor{blue}{0.8900} & \textcolor{blue}{29.88/0.8250}  \\ \cline{2-10}
& BTSRN~\cite{btsrn} & \textcolor{blue}{410K} & 165.2G & \textcolor{blue}{31.85}/$-$ & \textcolor{blue}{28.20}/$-$ & \textcolor{red}{27.47}/$-$ & \textcolor{red}{25.74}/$-$ & $-/-$  & $-/-$  \\ \cline{2-10} 
& \textcolor{red}{CARN-M~\cite{carn}}     & 412K  & \textcolor{blue}{32.5G}  &  \textcolor{red}{31.92}/\textcolor{red}{0.8903} & \textcolor{red}{28.42/0.7762} & \textcolor{blue}{27.44}/\textcolor{red}{0.7304} & \textcolor{blue}{25.62}/\textcolor{red}{0.7694}  & $-/-$  & $-/-$  \\ \cline{2-10}
& \textcolor{blue}{\armSRDOT-XL ($f$=32, $m$=11)}    & \textcolor{red}{114.97K}  & \textcolor{red}{6.62G}  & 31.54/\textcolor{blue}{0.8866}  &  28.12/0.7712  & 27.31/0.7277  & 25.31/\textcolor{blue}{0.7604}  & \textcolor{blue}{29.04}/\textcolor{red}{0.8901} & \textcolor{red}{29.94/0.8266} \\ 
\hline
\end{tabular}
\label{resX4}
}
\end{table*}

\subsection{Quantitative Results}\vspace{-1mm}
Table~\ref{resX2} reports PSNR/SSIM for several networks on six datasets for $\times2$ SISR. For clarity, we have broken down the results into three regimes: (\textit{i}) Small networks with 25K parameters or less, (\textit{ii}) Medium networks with 25K-100K parameters, and (\textit{iii}) Large networks with more than 100K parameters. As evident, \armSR dominates in all three regimes. Specifically, in the small network category, \armSRDOT-M5 achieves significantly better PSNR/SSIM than FSRCNN~\cite{fsrcnn} while using a similar number of parameters (e.g., 13.52K \textit{vs.} 12.46K) and $\sim2\times$ fewer MACs (3.11G vs. 6.00G). Even our smallest CNN (\armSRDOT-M3) outperforms all prior models while using $2.6\times$ to $3\times$ fewer MACs. Since our main comparison is against tiny CNNs like FSRCNN, we have also trained FSRCNN on the same setup and reported its results in the table.

In the medium network regime, we compare against the most recent tiny super resolution network called TPSR~\cite{tpsr}. Note that, we have reported results for the TPSR-NoGAN setting since we have not focused on Generative Adversarial Networks (GANs) or any perceptual losses in this work. Clearly, \armSRDOT-M11 outperforms TPSR-NoGAN while requiring $2.2\times$ fewer parameters and MACs. Note that, some of the baselines such as TPSR-NoGAN~\cite{tpsr} and MOREMNAS~\cite{moremnas} were found using advanced NAS techniques and our (manually designed) \armSR still significantly outperforms them.

For the large network category, we clearly see that our \armSRDOT-XL network either beats or comes close to much larger and highly accurate networks like CARN-M~\cite{carn} (\armSR uses $3.75\times$ fewer MACs) and BTSRN~\cite{btsrn} (\armSR uses $8.55\times$ fewer MACs). Most interestingly, our medium-range network (\armSRDOT-M11) actually achieves very similar or better PSNR than the VDSR network~\cite{vdsr}, which has $\bm{97\times}$ more MACs than \armSRDOT-M11.

Similar results are obtained for $\times4$ SISR. Recall that, we did not add multiple convolution layers in the upsampling block for \armSRDOT. This leads to even bigger savings in MACs for our proposed network. Table~\ref{resX4} shows the results for small, medium, and large categories. \armSRDOT-M5 now achieves better PSNR/SSIM than FSRCNN~\cite{fsrcnn} with $4.4\times$ fewer MACs. In the medium regime, \armSRDOT-M11 either outperforms or comes very close to TPSR-NoGAN~\cite{tpsr} while needing nearly $2\times$ fewer MACs. In the large network category, \armSRDOT-XL achieves similar or better image quality than LapSRN~\cite{lapsrn} while using $22.5\times$ fewer MACs. Finally, PSNR/SSIM of \armSRDOT-M11, again, comes very close to VDSR~\cite{vdsr}. \armSRDOT-M11 requires $\bm{331\times}$ fewer MACs than VDSR. \textit{As a result, our \armSRDOT-M11 network achieves VDSR-level performance even though it has nearly the same number of MACs as FSRCNN for $\times2$ SISR and has $2.5\times$ fewer MACs than FSRCNN for $\times4$ SISR.} Hence, \armSR significantly outperforms state-of-the-art CNNs in image quality and compute costs.

For $\times4$ SISR (large regime), there is still room for improvement: \armSRDOT-XL is nearly 0.4dB away from large CNNs like CARN-M~\cite{carn} and BTSRN~\cite{btsrn} for datasets like Urban100. This gap can potentially be filled using more channels ($f$) or extra upsampling convolutions like in prior art. This is left as a future work.

\vspace{-2mm}
\paragraph{Learned Perceptual Image Patch Similarity (LPIPS)}
LPIPS~\cite{lpips} is a perceptual image quality metric~\cite{imgMet} that has been widely used in SISR~\cite{tpsr}. A lower LPIPS value indicates that the image is more perceptually similar to the ground truth. Table~\ref{tab:lpips} presents the average LPIPS score for FSRCNN, \armSRDOT-M5, and \armSRDOT-M11 ($\times2$ SISR) on DIV2K and Urban100 datasets. Clearly, with lower or similar MACs, SESR results in lower LPIPS (i.e., better perceptual quality).
\begin{table}[]
\caption{Average LPIPS for Perceptual Image Quality ($\times2$ SISR)}
\centering
\scalebox{1.0}{
\begin{tabular}{|l||c|c|c|}
\hline
Model   & MACs & DIV2K & Urban100 \\ \hline \hline
FSRCNN   & 6.00G  & 0.107  & 0.110  \\ \hline
\armSRDOT-M5  & \textbf{3.11G}  & 0.103  & 0.094  \\ \hline
\armSRDOT-M11 & 6.30G  & \textbf{0.099}  & \textbf{0.085}  \\ \hline
\end{tabular}
\label{tab:lpips}
}\vspace{-4mm}
\end{table}

\begin{figure*}[tb]
\centering
\includegraphics[width=0.85\textwidth]{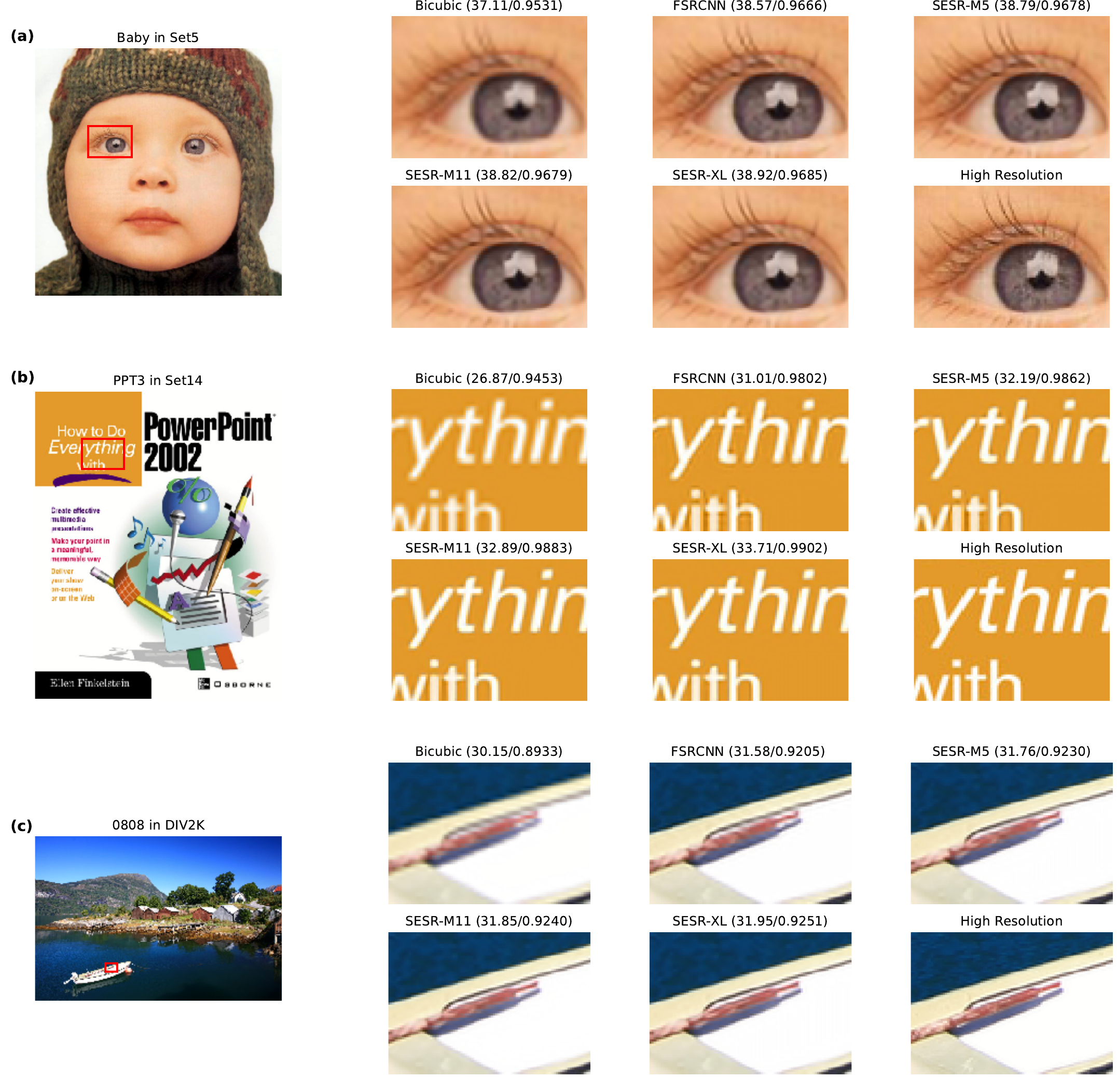}\vspace{-3mm}
	\caption{Qualitative comparison on $\times2$ SISR. \armSRDOT-M5 shows significantly better image quality while needing $2\times$ fewer MACs than FSRCNN. \armSRDOT-M11, which has similar MACs as FSRCNN, yields even better results. Numbers in parenthesis indicate PSNR/SSIM.\vspace{-2mm}}
\label{fig4}
\end{figure*}

\subsection{Qualitative Evaluation}\vspace{-1mm}
Fig.~\ref{fig4} below and Fig.~\ref{fig5} (in Appendix~\ref{AppMain}) show the image quality of various CNNs on $\times2$ and $\times4$ SISR, respectively. Since our focus is explicitly on highly efficient networks, we have compared the image quality of small- or medium-regime \armSR against other small networks like FSRCNN~\cite{fsrcnn}\footnote{Medium \armSRDOT-M11 is considered since it needs either similar (for $\times2$ SISR) or even fewer (for $\times4$ SISR) MACs than FSRCNN.}. As a reference for other high-quality models, we have provided the image for \armSRDOT-XL. Clearly, \armSRDOT-M5 outperforms FSRCNN (e.g., significantly sharper edges and less unwanted halo). \armSRDOT-M11 network performs even better than FSRCNN in all cases. Same holds for \armSRDOT-XL network. More qualitative results for $\times2$ and $\times4$ SISR are shown in Appendix~\ref{appMoreRes} (see Fig.~\ref{fig6}, Fig.~\ref{fig7}). Therefore, \armSR achieves significantly better image quality than other CNNs in similar compute regime.

\subsection{Training Efficiency of \armSR}\vspace{-1mm}
As described in Section~\ref{sec:etm}, the training cost is greatly reduced with our proposed training method for \armSRDOT. To demonstrate this, we trained three versions of \armSRDOT-M11:\vspace{-2mm}
\begin{itemize}
    \item Model A: VGG-like CNN without any linear blocks but with long residuals present, i.e., we directly train the model in Fig.~\ref{fig2}(d) without any linear blocks.
    \item Model B: Fully expanded \armSRDOT-M11 in Fig.~\ref{fig2}(a): forward pass happens through a fully expanded linear block. This is a traditional method to train linearly overparameterized networks~\cite{ExpandNets2020}.
    \item Model C: SESR-M11 trained with our efficient training method for linear blocks (see Section~\ref{sec:etm} and Fig.~\ref{effTr}).\vspace{-2mm}
\end{itemize}
We report the training time for each model (1 epoch, averaged over 10 epochs) on a single NVIDIA V100 GPU in Table~\ref{tab:effTr}. Clearly, our efficient training method trains linear blocks with minimal overhead compared to the completely collapsed VGG-like CNN and is significantly more efficient than traditional training of overparameterized networks.
\begin{table}[]
\caption{Training Time Efficiency of \armSR}
\centering
\scalebox{0.8}{
\begin{tabular}{|l|c|c|c|}
\hline 
\armSRDOT-M11 variants: & Model A & Model B & Model C \\ \hline
Time to train 1 epoch (s) & 30.6  & 116  & 34.9  \\ \hline
\end{tabular}
\label{tab:effTr}
}\vspace{-4mm}
\end{table}

\subsection{Comparison with Overparameterization Schemes}\label{sec:comparisonOver}\vspace{-1mm}
We now demonstrate how \armSR outperforms state-of-the-art overparameterization methods: ExpandNets and RepVGG. We modify the \armSRDOT-M11 network (with total 13 layers: eleven $3\times 3$ and two $5\times 5$) for these experiments. 

\vspace{-3mm}
\paragraph{Comparison against ExpandNets.}
In support of theory in Section~\ref{sec:theory}, we found that short residuals are essential for training overparameterized networks for SISR tasks. Specifically, we trained the \armSRDOT-M11 model using the exact same setup (e.g., learning rate, optimizer, etc.), but without the short residuals over $3\times 3$ linear blocks (\textit{i.e.}, long blue and black residuals in Fig.~\ref{fig2}(a) still exist). That is, this network is trained exactly using the procedure described by ExpandNets~\cite{ExpandNets2020}. This model quickly converged to 33.65dB DIV2K validation PSNR and did not improve further. In contrast, \armSRDOT-M11 achieves 35.45dB. Hence, the short residuals in our work are critical to obtain high accuracy on SISR tasks. This also suggests that the trainability of ExpandNet-type networks can suffer if short residuals are not used due to the vanishing gradient problem arising from increasing depth. Again, standard residual connections (like ResNets) over ExpandNets lead to significant memory transactions that result in heavy performance overhead on constrained hardware. This is because feature map sizes in SISR are huge (e.g., $1920\times 1080$) and residual connections require storing an extra feature map. Hence, collapsing the residuals is extremely important.

\vspace{-2mm}
\paragraph{Comparison against RepVGG.} We now train a RepVGG block (i.e., we overparameterized the $k\times k$ convolution with a $1\times1$ branch along with a short residual connection). This network achieved around 35.35dB which is lower than the 35.45dB achieved by \armSRDOT-M11 network. We then directly trained a completely collapsed \armSRDOT-M11 model (i.e., the VGG-like network with two long residuals shown in Fig.~\ref{fig2}(d)): This network achieved 35.34dB PSNR. Therefore, \textit{RepVGG performs nearly the same as VGG-like networks when the models are not sufficiently deep, which is exactly what our theory predicted in Section~\ref{sec:theoryRep}}. The proposed SESR performs the best out of existing methods. 

\subsection{Ablation Study: Residuals and PReLU vs. ReLU}\label{sec:preluAbl}\vspace{-1mm}
Next, we trained \armSRDOT-M11 with both long and short residuals shown in Fig.~\ref{fig2}(a) but without the linear blocks (\textit{i.e.}, only single $k\times k$ convolutions are used throughout the network with short residuals). This model converged to 35.25dB on DIV2K (compared to 35.45dB for \armSRDOT-M11). Thus, short residuals alone are not sufficient without linear blocks. Note that, the PSNR increase of even with 0.1 or 0.2dB over existing models is significant  (and often visually perceivable) since the model size is so small for our networks (standard deviation for all CNNs is very small $\sim0.02$dB).

Finally, we replace all PReLU activations with ReLU activations in \armSRDOT-M11 network and also remove the long input-to-output residual (see long, black residual in Fig.~\ref{fig2}(a)). Both of these changes can further improve hardware efficiency of \armSR and this model will be used in the next section to show (estimated) hardware performance results on a commercial \armNPU mobile-NPU. This CNN loses only about 0.1dB PSNR on DIV2K dataset. Hence, this variant of \armSR still significantly outperforms other similar sized networks like FSRCNN. Replacing PReLU with ReLU in FSRCNN also incurs a similar loss of PSNR.

\subsection{NPU Hardware Performance Estimation Results}\vspace{-1mm}
The system architecture can be summarized as follows:
CPU $\leftrightarrow$ DRAM $\leftrightarrow$ (DMA) $\leftrightarrow$ NPU SRAM $\leftrightarrow$ NPU MAC array. We use the performance estimator for \armNPU NPU for different models running 1080p to 4K ($\times2$) and 1080p to 8K ($\times4$) SISR. Table~\ref{hardwareRes} first shows MACs, DRAM Usage (to account for data movement between external and on-chip memories), Runtime and FPS for FSRCNN~\cite{fsrcnn} and \armSRDOT-M5\footnote{For hardware efficiency, we replace PReLU with ReLU in both \armSRDOT-M5 and FSRCNN, and also removed the input-to-output residual in \armSRDOT-M5. Both networks lose similar PSNR (0.1dB).} when converting a 1080p image to 4K resolution. As evident, even though \armSRDOT-M5 has $2\times$ fewer MACs than FSRCNN, the runtime is improved by $6.15\times$. This is because the hardware performance is guided not just by MACs but also the memory bandwidth\footnote{If data is not available to MAC units, they cannot compute. Hence, both memory usage and MACs are important for efficiency.}. The memory bandwidth for SISR is heavily dependent on the activation sizes. For FSRCNN, the size of the largest activation tensor is $H\times W\times 56$, whereas for \armSRDOT-M5, it is $H\times W\times 16$, where $H\times W$ are the dimensions of low-resolution input. That is, \armSRDOT-M5's largest tensor is $3.5\times$ smaller than that of FSRCNN and thus the DRAM use is correspondingly $2\times$ smaller in \armSRDOT-M5 than FSRCNN. This results in an overall $6.15\times$ better runtime. This shows the challenges of running real-world SISR on constrained devices and how \armSR significantly outperforms FSRCNN.
\begin{table}[]
\caption{Performance Estimation for \armNPU NPU}
\centering
\scalebox{0.7}{
\begin{tabular}{|l|c|c|c|c|} \hline
Model and& MACs & DRAM & Runtime (ms) & Improvement \\
Resolution &  &  Use (MB) & /FPS & (Runtime) \\ \hline \hline
\begin{tabular}[c]{@{}l@{}}FSRCNN ($\times2$)\\ 1080p$\rightarrow$4K\end{tabular}    & $54$G &  $564.11$ & $167.38/5.97$ & $1\times$ \\ \hline
\begin{tabular}[c]{@{}l@{}}\armSRDOT-M5 ($\times2$)\\ 1080p$\rightarrow$4K\end{tabular}   & $28$G  & $282.03$  &  $27.22/36.73$ & $\bm{6.15\times}$ \\ \hline 
\begin{tabular}[c]{@{}l@{}}\armSRDOT-M5 (Tiled, $\times2$)\\  400$\times$300$\rightarrow$800$\times$600\end{tabular} &  $1.62$G &  $6.46$ &  $1.26/792.38$   & $-$ \\ \hline \hline
\begin{tabular}[c]{@{}l@{}}\armSRDOT-M5 ($\times4$)\\ 1080p$\rightarrow$8K\end{tabular}   & $38$G  & $389.86$  &  $45.09/22.17$ & $\bm{>3.7\times}$ \\ \hline 
\begin{tabular}[c]{@{}l@{}}\armSRDOT-M5 (Tiled, $\times4$)\\  400$\times$300$\rightarrow$1600$\times$1200\end{tabular} &  $2.19$G &  $9.84$ &  $2.12/471.69$   & $-$ \\ \hline 
\end{tabular}
\label{hardwareRes}
}\vspace{-2mm}
\end{table}

\vspace{-2mm}
\paragraph{Further optimizations to get up to $\bm{8\times}$ better runtime.}
As mentioned, DRAM usage for SISR application is naturally very high due to large input images (e.g., a 1080p input has $1920\times 1080$ dimensions). To further accelerate the inference, the input can be broken down into tiles so that the DRAM traffic is minimized. As a proof-of-concept of this optimization, we divide a 1080p image into tiles of $400\times 300$ and perform a $400\times300\rightarrow800\times600$ SISR. The performance numbers for this tile are given in Table~\ref{hardwareRes}. Clearly, we need to do this inference at least $(1920/400)\times(1080/300)=17.28$ times to cover the entire input image. Hence, total inference time is given by (Performance for one tile $\times 17.28$) which comes out to about $21.77$ms or $\approx 46$FPS (nearly $8\times$ faster than FSRCNN: 6FPS vs. 46FPS). Note that, these are only approximate calculations. In the real-world, there will be (\textit{i}) boundary overhead when tiling image to maintain the functional correctness, and (\textit{ii}) other software overheads. However, since \armSRDOT-M5 network is not very deep, these overheads are not significant. This also brings us a little closer to 60FPS on a mobile-NPU when performing 1080p to 4K SISR.

Recall that, for $\times4$ super resolution, \armSR scales up better than FSRCNN in MACs. Hence, FSRCNN will achieve much less than 6FPS for 1080p to 8K SISR. In contrast, Table~\ref{hardwareRes} shows that \armSRDOT-M5 achieves 22FPS which is at least $3.7\times$ better than even $\times2$ (1080p to 4K) FSRCNN's 6FPS. Therefore, \armSR will achieve significantly better performance than FSRCNN for 1080p to 8K SISR. Note that, we have estimated the final depth-to-space for our $\times4$ network using [1080p to 4K] and [4K to 8K] (both using $\times2$ SISR), instead of a one-shot $\times4$ depth-to-space from 1080p to 8K. Hence, these numbers are still somewhat pessimistic and may be improved further using a one shot $\times4$ depth-to-space operation. Finally, similar to the $\times2$ SISR case above, with tiling, the 22FPS can be improved up to 27FPS (a runtime of $2.12\times (1920/400)\times(1080/300)$ leads to 27FPS; see $\times4$ tiling results in Table~\ref{hardwareRes}). Thus, \armSR enables 1080p to 4K and 1080p to 8K super resolution with significantly faster frame rates on commercial mobile-NPUs.

\vspace{-2mm}
\paragraph{Additional improvement in inference time using even-sized and asymmetric kernels.}
For a $200\times200\rightarrow400\times400$ SISR task on DIV2K dataset, we found that the NAS-guided network produced by targeting NPU latency reduced the inference time by $15\%$ in comparison to the \armSRDOT-M5 network while exactly matching its accuracy. This is primarily attributed to the use of smaller sized kernels for the first and last linear blocks, smaller even-sized $2\times2$ kernels for intermediate linear blocks $1$, $2$ and $4$ and smaller asymmetric $2\times1$, $3\times2$ and $2\times3$ kernels for intermediate linear blocks $3$, $5$, $6$, and $7$ in the NAS-guided \armSR network as shown in Fig.~\ref{nas_sesr_same_accuracy}(b) (in Appendix~\ref{appNAS}). This essentially demonstrates the efficacy of even-sized and asymmetric kernels in boosting the performance of further optimized \armSR networks on a commercial NPU hardware.

\subsection{Open Source NPU Performance Estimation}\vspace{-1mm}
To enable open source performance estimation for a commercial NPU, we next consider Arm Ethos-U55 which is a tiny micro-NPU (0.5 TOP/s) that can accompany microcontrollers like Cortex-M-based systems~\cite{u55}. The performance estimator for Ethos-U55 is publicly available and is called Vela\footnote{Vela is available at: \url{https://git.mlplatform.org/ml/ethos-u/ethos-u-vela.git/about/}}. To estimate performance on this micro-NPU, we run FSRCNN~\cite{fsrcnn} and \armNPUDOT-M5 models through Vela version 3.2. For $\times2$ SISR (360p to 720p), FSRCNN achieves only 2.71 FPS (369.66ms latency), whereas our proposed \armSRDOT-M5 achieves 22.67FPS (44.11ms latency). Therefore, \armSRDOT-M5 improves FPS over FSRCNN by more than $8\times$ on the Ethos-U55. Hence, the NPU results can now be verified using this open source tool. 

\subsection{Real Hardware Performance Results}\vspace{-1mm}
Finally, we deploy SESR-M5 and FSRCNN on a real mobile device containing four Arm Cortex A77 CPUs and Arm Mali G78 GPU. We considered the following SISR tasks: (a) 360p to 720p ($\times2$), and (b) 180p to 720p ($\times4$). Moreover, we used the optimized CPU-XNNPack and TFLITE GPU-delegate libraries to run models on the CPU and GPU, respectively. As evident from Table~\ref{cpugpu}, SESR-M5 is significantly ($\bm{1.5\times}$-$\bm{2\times}$) faster on real mobile CPU and GPU than FSRCNN while improving image quality.
\begin{table}[]
\caption{Latency Results on a Real Arm CPU and GPU}
\scalebox{0.95}{
\centering
\begin{tabular}{|l||cc||cc|}
\hline
\multirow{3}{*}{Model} & \multicolumn{2}{c||}{\multirow{2}{*}{\begin{tabular}[c]{@{}c@{}}Latency (ms)\\ $\times2$ (360p to 720p)\end{tabular}}} & \multicolumn{2}{c|}{\multirow{2}{*}{\begin{tabular}[c]{@{}c@{}}Latency (ms)\\ $\times4$ (180p to 720p)\end{tabular}}} \\
                   & \multicolumn{2}{c||}{}                                                               & \multicolumn{2}{c|}{}                                                               \\ \cline{2-5} 
                   & \multicolumn{1}{c|}{CPU}                               & GPU                             & \multicolumn{1}{c|}{CPU}                              & GPU                             \\ \hline \hline
FSRCNN             & \multicolumn{1}{c|}{111.8}                               &           32.93                    & \multicolumn{1}{c|}{31.51}                               &   19.95                            \\ \hline
\armSRDOT-M5       & \multicolumn{1}{c|}{60.39}                               &      20.38                         &  \multicolumn{1}{c|}{20.56}                               &   10.14                            \\ \hline
Improvement        & \multicolumn{1}{c|}{$\bm{1.85\times}$}                               &               $\bm{1.61\times}$                & \multicolumn{1}{c|}{$\bm{1.53\times}$}                               &                 $\bm{1.96\times}$              \\ \hline
\end{tabular}
}\label{cpugpu}\vspace{-2mm}
\end{table}

\section{Conclusion}\label{sec:conclusion}\vspace{-2mm}
In this paper, we have proposed \armSR networks that establish a new state-of-the-art for efficient super resolution. Our proposed models are based on collapsible linear blocks, a linear overparameterization technique. With a theoretical analysis, we have discovered that recent overparameterization techniques like RepVGG do not present any advantages over non-overparameterized VGG-like CNNs when the networks are not too deep. The proposed \armSR alleviates these theoretical limitations.  Detailed experiments across six datasets demonstrate that \armSR achieves similar or better image quality than state-of-the-art CNNs while using $\bm{2\times}$ to $\bm{330\times}$ fewer MACs. This enables \armSR to efficiently perform $\times2$ (1080p to 4K) and $\times4$ SISR (1080p to 8K) on resource constrained devices. To this end, we estimate hardware performance for 1080p to 4K ($\times2$) and 1080p to 8K ($\times4$) SISR on a small mobile-NPU. We found that \armSR achieves $6\times$-$8\times$ higher FPS than prior art on the \armNPU NPU. Further performance gains are obtained using a proof-of-concept NAS on \armSR search space. Finally, we have shown real performance improvements ($1.5\times$-$2\times$) on Arm CPU and GPU by deploying \armSR on a mobile device.

\bibliography{example_paper}
\bibliographystyle{mlsys2022}


\clearpage

\appendix
\section*{Supplementary Information:\\ Collapsible Linear Blocks for Super-Efficient Super Resolution}
\section{$\bm{\times4}$ Results from Main Text}\label{AppMain}
\begin{figure*}[h]
\centering
\includegraphics[width=0.87\textwidth]{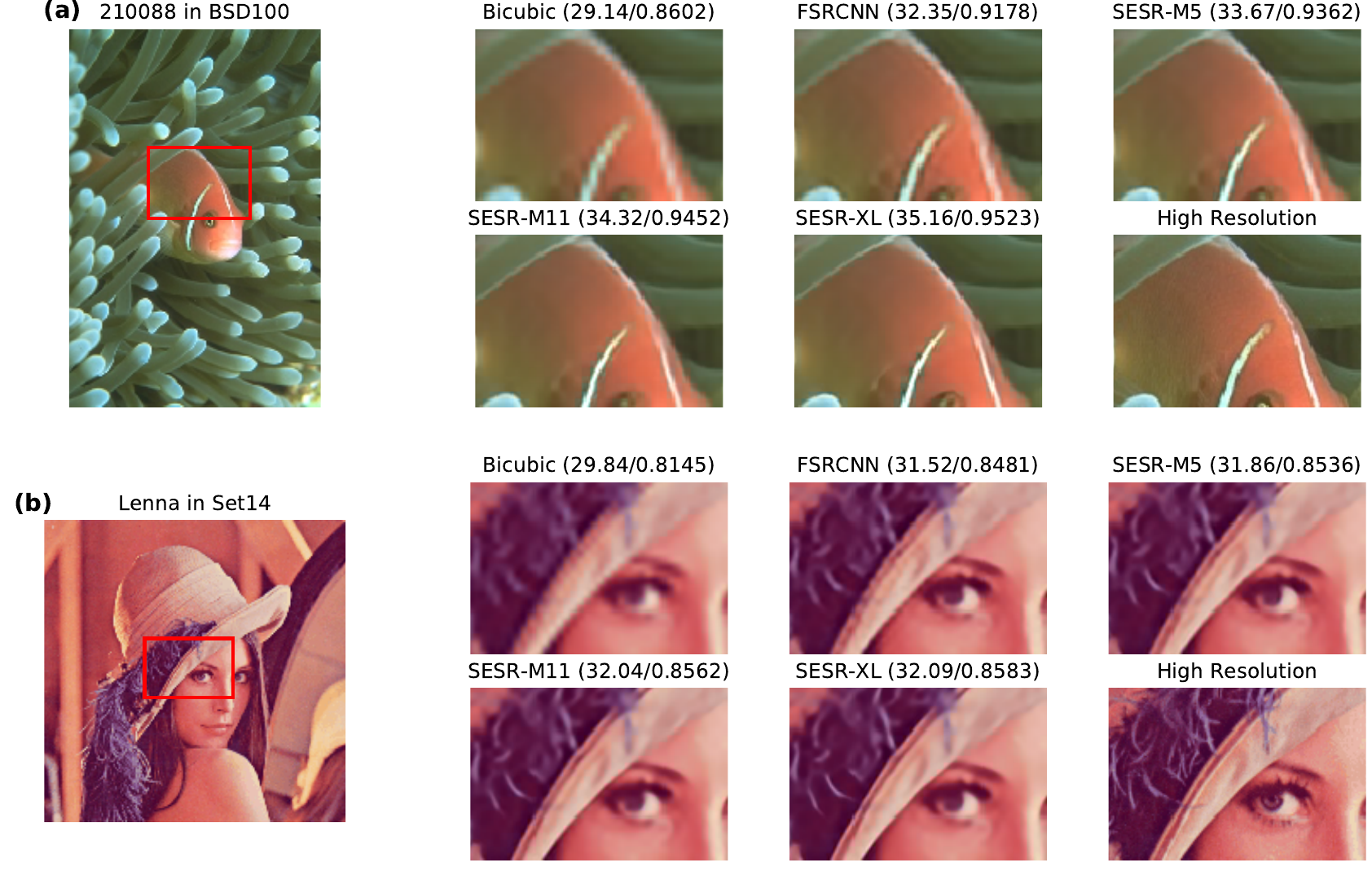}\vspace{-2mm}
	\caption{Qualitative comparison on $\times4$ SISR. Both \armSRDOT-M5 and \armSRDOT-M11 require significantly fewer MACs than FSRCNN and yield better image quality (e.g., better edges, no unwanted halo, \textit{etc.}). Numbers in parenthesis indicate PSNR/SSIM.}
\label{fig5}
\end{figure*}

Fig.~\ref{fig5} shows the $\times4$ results from main text. Please see the next page. 

\section{Additional Qualitative Results}\label{appMoreRes}
\begin{figure*}[!tbh]
\centering
\includegraphics[width=0.9\textwidth]{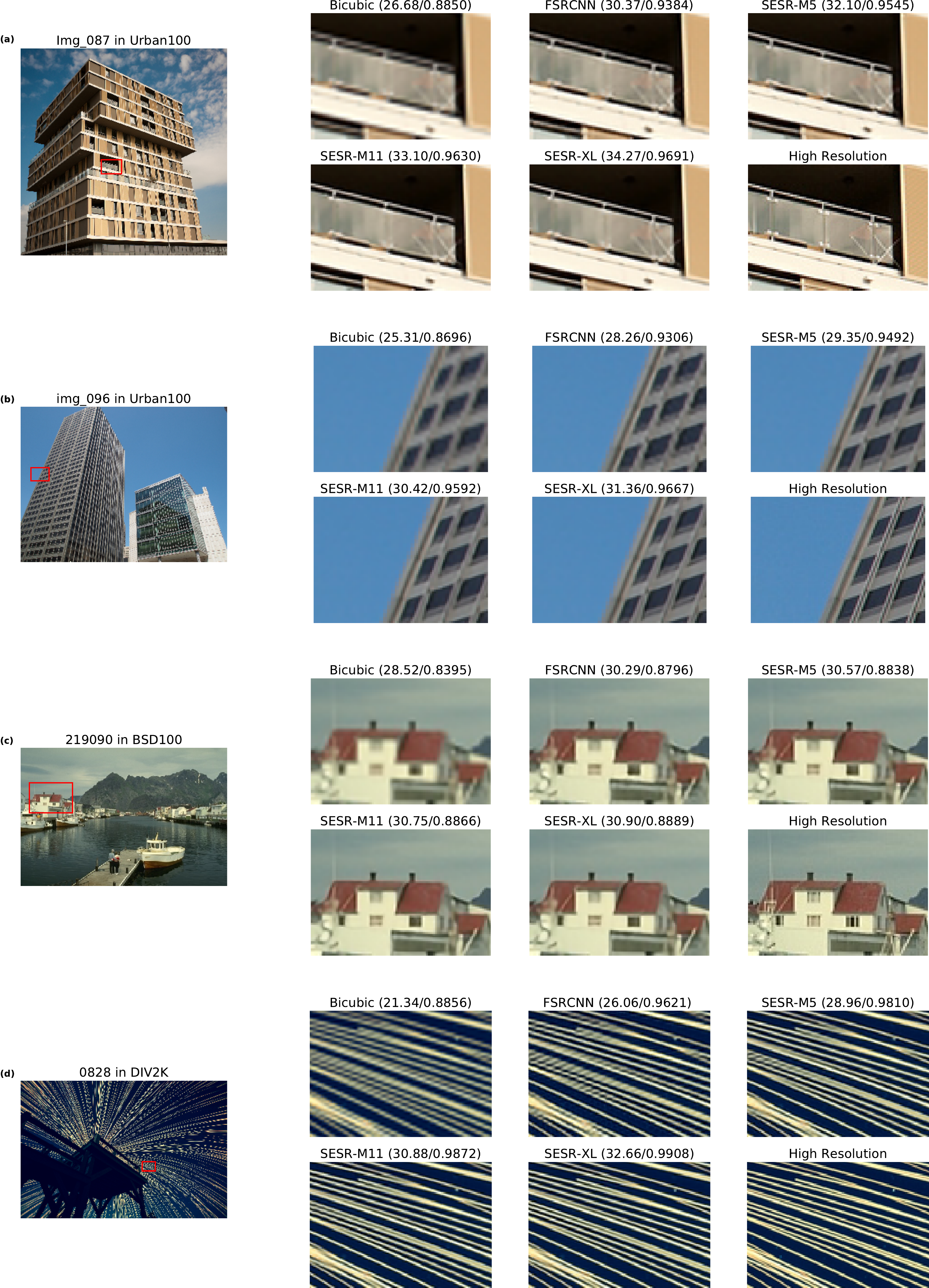}\vspace{-1mm}
	\caption{Additional Results: Qualitative comparison on $\times2$ SISR. \armSRDOT-M5 shows much better image quality while needing $2\times$ fewer MACs than FSRCNN. \armSRDOT-M11 (similar MACs as FSRCNN) yields even better results. Numbers in parenthesis indicate PSNR/SSIM.\vspace{-2mm}}
\label{fig6}
\end{figure*}

\begin{figure*}[tb]
\centering
\includegraphics[width=0.9\textwidth]{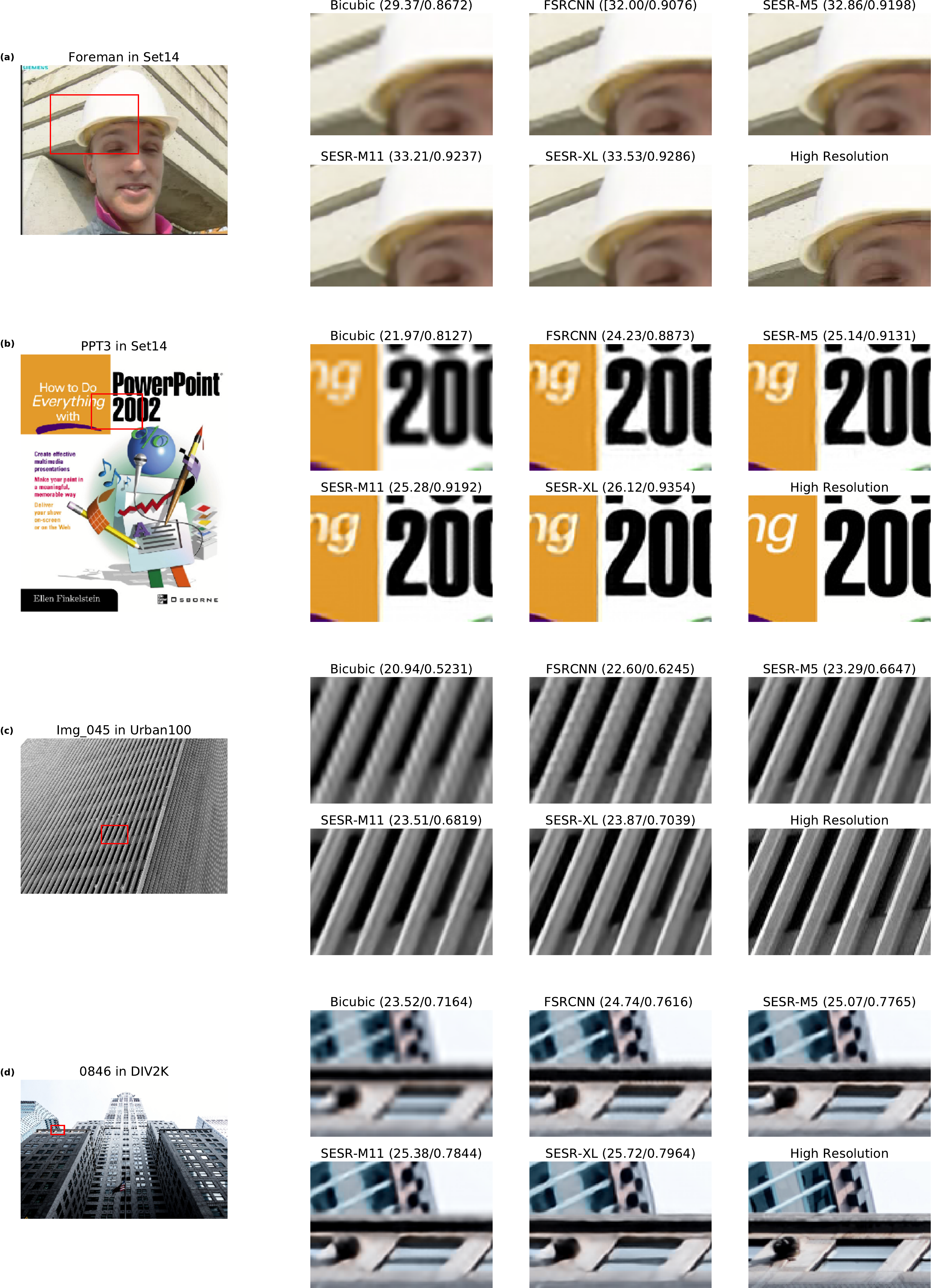}\vspace{-1mm}
	\caption{Additional Results: Qualitative comparison on $\times4$ SISR. Both \armSRDOT-M5 and \armSRDOT-M11 require significantly fewer MACs than FSRCNN and yield better image quality (e.g., better edges, no unwanted halo, \textit{etc.}). Numbers in parenthesis indicate PSNR/SSIM.\vspace{-2mm}}
\label{fig7}
\end{figure*}

Results for both $\times2$ and $\times4$ SISR are shown in Fig.~\ref{fig6} and Fig.~\ref{fig7}. Please see the next few pages.

\section{NAS Searched Models}\label{appNAS}
\begin{figure*}[!tbh]
\centering
\includegraphics[width=0.3\textwidth]{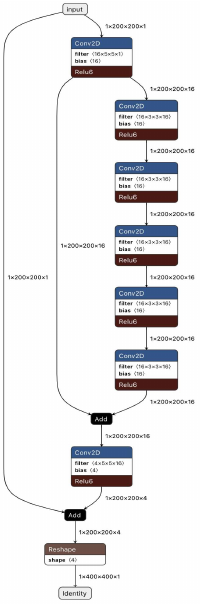}\vspace{-1mm}
\includegraphics[width=0.3\textwidth]{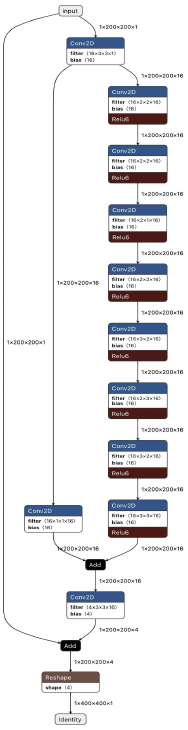}\vspace{-1mm}
\includegraphics[width=0.3\textwidth]{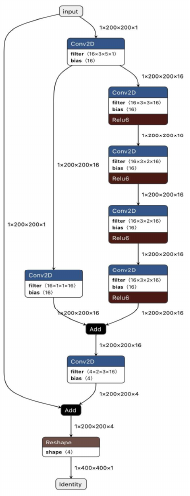}\vspace{-1mm}
	\caption{(a) Left: Manually designed \armSRDOT-M5 network with two $5\times 5$ and five intermediate $3\times 3$ linear blocks, (b) Center: NAS-guided \armSR network with two $3\times3$ and eight intermediate linear blocks. The eight intermediate linear blocks consist of $2\times2$, $2\times1$, $2\times3$, $3\times2$ and $3\times3$ kernels as opposed to only $3\times3$ kernels in the five intermediate blocks of \armSRDOT-M5. NAS-guided \armSR network observes $15\%$ reduction in inference time in comparison to \armSRDOT-M5 while offering same PSNR, (c) Right: NAS-guided \armSR network produced by targeting $50\%$ latency of the \armSRDOT-M5. It achieves same PSNR to that of the \armSRDOT-M3 while requiring less inference time than \armSRDOT-M3.\vspace{-2mm}}
	
\label{nas_sesr_same_accuracy}
\end{figure*}

Fig.~\ref{nas_sesr_same_accuracy} shows the NAS searched models.

\end{document}